\documentclass[aps,onecolumn,preprint,showpacs]{revtex4}
\usepackage{epsfig}
\usepackage{amsmath}
\usepackage{amssymb}
\usepackage{amsfonts}
\usepackage{enumerate}

\begin{document}
%\draft
\title{Transverse spin relaxation time in organic molecules: A possible 
platform for fault tolerant room temperature quantum computing}
\author{B. Kanchibotla, S. Pramanik and S. 
Bandyopadhyay$\thanks{Corresponding author. E-mail: sbandy@vcu.edu.}$}
\affiliation{Department of Electrical and Computer Engineering, Virginia Commonwealth 
University, Richmond, VA 23284, USA}
\author{M. Cahay}

\affiliation{Department of Electrical and Computer Engineering, University of 
Cincinnati, OH 45221, USA}

\begin{abstract}

We report measurement of the ensemble averaged transverse spin relaxation 
time $(T_2^*)$ in bulk and few molecules of the organic semiconductor  
$tris(8-hydroxyquinolinolato~
aluminum)$ or 
$Alq_3$. This system exhibits two characteristic $T_2^*$ times, the longer 
of which is temperature-independent and the shorter is temperature-dependent, 
indicating that the latter is most likely limited by 
spin-phonon interaction. 
Based on the measured data, we infer that the single particle $T_2$ time is long enough to meet
Knill's criterion for fault tolerant quantum computing, even at {\it
room temperature}. $Alq_3$ is also an optically active organic and 
 we propose a simple optical scheme for spin qubit read out.
Moreover, we found that the temperature-dependent $T_2^{*}$ time is considerably 
shorter in bulk $Alq_3$ powder than in few molecules confined in 1-2 nm 
sized cavities, which is suggestive of  a new type of ``phonon bottleneck effect''. This is 
very intriguing for organic molecules where carriers are always localized over 
individual molecules but the phonons are delocalized.
\end{abstract}
\maketitle

The $\pi$-conjugated organic semiconductor $Alq_{3}$ exhibits exceptionally long 
longitudinal spin relaxation time $T_{1}$ (approaching
1 second at 100 K) because of weak spin-orbit interactions \cite{pramanik}. 
That bodes well for classical spin based devices like Spin Enhanced Organic Light Emitting Diodes
\cite{salis} or {\it classical} spin based computing paradigms such as Single Spin Logic 
\cite{miller,super} where a long $T_{1}$ time reduces the probability of bit errors 
caused by unwanted spin flips. 
In {\it quantum} computing paradigms \cite{bandyopadhyay,loss,kane,phys,calarco,popescu}, 
the bit error probability depends on the transverse spin relaxation time $T_{2}$ rather than $T_{1}$. 
The probability of a spin based qubit to decohere during a qubit operation that lasts for a time duration $T$
is roughly $1-e^{\frac{-T}{T_2}}$. Knill has shown that fault tolerant quantum computing becomes possible 
if this probability is less than 3\% \cite{knill}, i.e., if 
$T_2/T~>$ 33. 

Two recent results have inspired us to look towards the $Alq_{3}$ molecule as a potential candidate 
for fault-tolerant spin based quantum computing. The first is the demonstration that it exhibits a long $T_{1}$ time 
\cite{pramanik}. This results from weak spin orbit interactions which could also make the $T_{2}$ time 
long enough to allow fault tolerant 
computing. 
Second, some organic molecules can be efficient quantum processors with high gate fidelity \cite{lehman}. 
These two factors, taken together, raise the hope that $Alq_3$ might be a preferred
platform for spin based quantum computing. This molecule also has spin-sensitive optical 
transitions that can be gainfully employed for spin (qubit) read out. That makes it even more attractive.

Unfortunately, it is very difficult to measure the single particle $T_{2}$ time directly in any system 
(including $Alq_{3}$ molecules) since it requires complicated spin echo sequences. 
Therefore, we have measured the ensemble averaged $T_2^{*}$ time instead, 
since it can be ascertained easily from the line width of electron spin resonance spectrum. 
This time however is orders of magnitude shorter than the actual $T_{2}$ time 
of an isolated spin because of additional decoherence caused by interactions between 
multiple spins in an ensemble \cite{desousa,hu}. It is particularly true of 
organics where spin-spin interaction is considered to be the major mechanism for spin decoherence
\cite{sanvito}. Consequently, bulk samples (where numerous spins interact with each other) 
should behave differently from one or few molecules containing fewer interacting spins. 
In the rest of the paper, we will designate the $T_2^{*}$ times of bulk and few-molecule samples as 
$T_2^{b}$ and $T_2^{f}$, respectively.
We have found that they are discernibly different.

In order to prepare samples containing one or few molecules, we followed the approach in ref. \cite{huang}. 
We first produced a porous alumina film with 10-nm pores by anodizing an aluminum foil in 15\% sulfuric acid
\cite{nanotechnology}. A two-step anodizing process was employed to improve the 
regimentation of the pores \cite{masuda}. These porous films were then soaked in 1, 2-dichloroethane 
$(C_2H_4Cl_2)$ solution of $Alq_{3}$ for over 24 hours to impregnate the pores with $Alq_{3}$ molecules. 
The films were subsequently washed several times in pure $C_2H_4Cl_2$ to remove excess $Alq_{3}$. 
There are cracks of size 1-2 nm in the anodic alumina film produced in sulfuric acid \cite{huang,macdonald,ono}.
 Ref. 
\cite{huang} claims that when the anodic alumina film is soaked in $Alq_{3}$ solution, 
$Alq_{3}$ molecules of 0.8 nm size diffuse into the cracks and come to rest in nanovoids nestled 
within the cracks. Since the cracks are 1-2 nm wide, only  1-2 molecules of $Alq_3$ can reside 
in the nanovoids.  
Surplus molecules, not in the nanovoids, will be removed by repeated rinsing in $C_2H_4Cl_2$ \cite{huang}. 
$C_2H_4Cl_2$ completely dissolves out all the $Alq_3$ molecules, except those in the nanovoids, 
because the $C_2H_4Cl_2$ molecule cannot easily diffuse through the 1-2 nm wide nanocracks
to reach the nanovoids. Therefore, after the repeated
rinsing procedure is complete, we are left with an ensemble of few-molecule clusters in the nanovoids.
The nanovoids are sufficiently far from each other that interaction between them is negligible
\cite{huang}. Therefore, if we use the fabrication technique of ref. \cite{huang}, we will be 
confining one or two isolated molecules in nanovoids and measuring their $T_2^{f}$ times. 
In contrast, the $T_2^{b}$ times are measured in bulk $Alq_{3}$ powder containing a very large number 
of interacting molecules.

The $T_2^{f}$ and $T_2^{b}$ times were measured using electron spin resonance (ESR) spectroscopy. 
In each run, 20 samples of area 25 mm$^2$ each were stacked. They together contain > 5$\times$10$^{11}$ pores
and even if each pore contains at least one molecule and each molecule contains at least one electron (extremely 
conservative estimate), we still have over 5$\times$10$^{11}$ electrons, which are more than adequate 
to provide a strong spin signal (our equipment can measure signals from 10$^9$ spins). 

It is well known that $Alq_{3}$ has two spin resonances corresponding to Land\'{e}\ g-factors of approximately 
2 and 4 \cite{mirea}. Ref. \cite{mirea} determined from the temperature dependence of the 
ESR intensity that the $g$ = 4 resonance is associated with localized spins in $Alq_{3}$ 
(perhaps attached to an impurity or defect site) while the $g$ = 2 resonance is associated with 
quasi free (delocalized) spins. From the measured line widths of these two resonances, 
we can estimate the  $T_2^{f}$ and $T_2^{b}$ times for each resonance individually using the standard formula
\begin{equation}
T_2^{f} or T_2^{b}=\frac{1}{r_e\left( g/2\right) \sqrt{3}\Delta B_{pp}}
\end{equation}
where $r_e$ is a constant = 1.76 x 10$^{7} (G-s)^{-1}$, $g$ is the Land\'e g-factor and $\Delta B_{pp}$ is 
the full-width-at-half-maximum of the ESR line shape (the "line width"). We checked that the line shape is 
almost strictly Lorentzian, so that the above formula can be applied with confidence \cite{rieger}.
Fig. 1 shows a typical magnetic field derivative of the ESR spectrum obtained at a temperature of 10 K 
corresponding to $g$ = 2 resonance. There are three curves  in this figure corresponding to the blank alumina host, 
bulk $Alq_{3}$ powder, and $Alq_{3}$ in 1-2 nm voids. The alumina host has an ESR peak at $g$ = 2 
(possibly due to oxygen vacancies) \cite{duh}, but it is much weaker than the resonance signals from 
$Alq_{3}$ and hence can be easily separated. Note that the g-factor of the isolated $Alq_{3}$ molecules 
in nanovoids is slightly larger than that of bulk powder since the resonance occurs at a slightly 
higher magnetic field. More importantly, the bulk powder has a broader line width than the few molecules 
confined in the nanovoids. This is a manifestation of the fact that stronger spin-spin interactions in the 
bulk powder reduce the effective $T_2^{*}$ time, i.e., $T_2^{b}$ $<$ $T_2^{f}$.

In Fig. 2, we plot the measured $T_2^{f}$ and $T_2^{b}$ times (associated with the resonance 
corresponding to $g$ = 2) as functions of temperature from 4.2 K to 300 K. The inequality 
$T_2^{b}$ $<$ $T_2^{f}$ is always satisfied except at one anomalous data point at 4.2 K.
There are two important points to note here. First, both $T_2^{f}$ and $T_2^{b}$ are relatively 
temperature independent over the entire range from 4.2 K to 300 K. This indicates that spin-phonon 
interactions do not play a significant role in spin dephasing.  Second, 
both $T_2^{f}$ and $T_2^{b}$ times are quite long, longer than 3 nanoseconds, 
even at room temperature. 

In Fig. 3, we plot the measured $T_2^{f}$ and $T_2^{b}$ times as functions of temperature 
corresponding to the $g$ = 4 resonance. The $T_2^{f}$ time is plotted from 4.2 K to 300 K, 
but the $T_2^{b}$ time in bulk powder can only be plotted up to a temperature of 100 K. 
Beyond that, the intensity of the ESR signal fades below the detection limit of our equipment. 
The features to note here are that: (1) $T_2^{f}$ and $T_2^{b}$ are no longer temperature independent 
unlike in the case of the $g$ = 2 resonance. $T_2^{f}$ decreases monotonically with increasing temperature 
and falls by a factor of 1.7 between 4.2 K and 300 K, 
(2) $T_2^{b}$ $<$ $T_2^{f}$ and the ratio $T_2^{f}/T_2^{b}$ 
decreases with increasing temperature. 
The maximum value of the ratio $T_2^{f}/T_2^{b}$ is 2.4, occurring at the 
lowest measurement temperature of 4.2 K, and (3) both $T_2^{f}$ and $T_2^{b}$ times 
are about an order of magnitude shorter for the $g$ = 4 resonance compared to the $g$ = 2 resonance. 

The strong temperature dependence of $T_2^{f}$ and $T_2^{b}$ tells us that for $g$ = 4 resonance, 
spin-phonon coupling plays the dominant role in spin dephasing instead of spin-spin interaction. 
The spin-phonon coupling is absent or significantly suppressed for the $g$ = 2 resonance, which is 
why $T_2^{f}$ and $T_2^{b}$ are an order of magnitude longer and also temperature independent for $g$ = 2. 
Ref. \cite{mirea} has ascribed the $g$ = 2 resonance to quasi free carrier spins in $Alq_{3}$ 
(whose wavefunctions are extended over an entire molecule) and $g$ = 4 
resonance to localized spins (whose wavefunctions are localized over an impurity atom). 
If that is the case, then it is likely that the localized spins and 
the delocalized spins will have very different couplings to phonons since their  wavefunctions 
are very different. 

An interesting question is why should $T_2^{f}$ be so much longer than $T_2^{b}$ for the $g$ = 4 resonance. 
The bulk has many more interacting spins than the few-molecule sample has, but if spin-spin interaction 
is overshadowed by spin-phonon coupling, then this should not make any difference. 
What could be causing this behavior is a new type of "phonon-bottleneck effect. 
For $g$ = 4 resonance, we know that the primary dephasing agents are phonons. 
So what makes the spin-phonon coupling so much stronger in bulk than in nanovoids? 
In bulk $Alq_{3}$ powder, the phonons are not confined and form a continuum. 
However, in isolated nanovoids (cavities) of $\sim$ 2 nm diameter, the phonons are confined so 
that only discrete phonon modes are allowed. Any dephasing transition will then have to emit 
or absorb a subset of these allowed phonon modes. This reduces the transition probability considerably 
since few phonons are available to satisfy the energy and momentum conservations for phonon 
emission and absorption. This is a new type of phonon bottleneck effect,
slightly different from the one discussed in ref."\cite{benisty}, which 
required carrier confinement more than phonon confinement. 
This new type of phonon bottleneck effect would explain why $T_2^{f} ~>$ $T_2^{b}$ 
when phonons are the primary dephasing agents. The bottleneck will be more severe at lower temperatures 
since fewer phonon modes will be 
occupied (Bose Einstein statistics), which is exactly what we observe. If this explanation is true, 
it will be the first observation of this effect in organic molecules. What makes it more intriguing is 
the fact that there is no "quantum confinement" effect on electrons since their wavefunction is at 
best extended over a single molecule which is only $\sim$ 0.8 nm in size, but 
the phonon modes are extended over many molecules and therefore do suffer 
quantum confinement if the confining space is a nanovoid of $\sim$ 2 nm in diameter. 
We raise the specter of phonon bottleneck only as a possibility, but cannot confirm it experimentally 
beyond all reasonable doubt since that would require showing progressive suppression of dephasing with 
decreasing nanovoid size, something that is experimentally not accessible. Nonetheless, we believe 
that there is a strong suggestion for the phonon bottleneck effect.

We conclude by discussing the suitability of $Alq_{3}$ molecules for quantum computing applications. 
For a single isolated spin in $Alq_{3}$, $T_2$ should be at least an order of magnitude longer than $T_2^{*}$ 
\cite{desousa,hu} particularly when spin-spin interaction is the major dephasing mechanism ($g$ = 2). 
Since we have measured that $T_2^{*}$ $\sim$ 3 nanosecond at nearly all temperatures between 4.2 K and
300 K for $g$ = 2 resonance, we expect that the single spin $T_2$ time will be at least 30 nanoseconds
over this entire temperature range.
Now, if Rabi oscillation is used for qubit operations such as rotation \cite{kane,phys}, then the time 
taken to effect a complete spin flip is $T$ = $ h/\left( 2g\mu_BB_{ac}\right)$ where $g$ is the Land\'e 
g-factor, $\mu_B$ is the Bohr magneton and $B_{ac}$ is the amplitude of the ac magnetic field inducing the 
Rabi oscillation. With $B_{ac}$ = 500 Gauss \cite{akhkalkatsi}, T = 0.35 nanoseconds. Therefore, 
the error probability  = $1- exp[-T/T_2]$ = 1.15\%. This is {\it less} than the Knill limit of 3\% for fault 
tolerant quantum computing, which is encouraging. 
We emphasize that $Alq_{3}$ does not have exceptionally long $T_2$ times, but it 
is still adequate for fault tolerant quantum computing. 
Nitrogen vacancy $NV^{-}$ in diamond exhibits a much longer $T_2$ time of several tens of 
$\mu$sec at room temperature \cite{colton}. 
However, quantum computing paradigms based on $NV^{-}$ require optical gating 
\cite{kilin,dutt} or cavity dark states \cite{bowers} since it would be nearly 
impossible to place an electrical gate on top of an atomic vacancy using any of 
the known fabrication methods. As a result, $NV^{-}$ computers are not truly scalable. 
In contrast, the spins in $Alq_{3}$ are not bound to specific atomic sites. Instead, they extend 
over molecules of size $\sim$ 1 nm, which allows 
electrical gating and therefore scalable renditions of quantum processors. 
Inorganic semiconductor qubit hosts, that will also allow electrical gating, typically have a shorter 
$T_2^{*}$ time than $Alq_{3}$ at room temperature \cite{ghosh}.
Therefore, the $Alq_3$ system
deserves due attention.

Finally, if an $Alq_{3}$ quantum dot were used as a host for a spin qubit, 
one would require a mechanism for reading the host spin (qubit read out). 
Fortunately, this can be achieved quite simply and elegantly. 
It is well known that only the singlet exciton recombines radiatively in $Alq_{3}$ and 
the triplet does not \cite{salis}. Thus, one needs to inject a spin polarized hole into an 
$Alq_{3}$ quantum dot that hosts a single electron in the LUMO level, 
from a p-type dilute magnetic semiconductor such as GaMnAs. 
The hole's spin will be known (majority spin in GaMnAs). If a photon is emitted from the 
$Alq_{3}$ quantum dot, then we will know that the electron's spin and the hole's spin are 
anti-parallel. Otherwise, they are parallel. This allows one to determine the electron's 
spin polarization in the $Alq_{3}$ dot (qubit read out). The optical read out mechanism requires a 
quantum dot photon detector to be integrated on top of the $Alq_{3}$ quantum dot hosting the spin.
This is not difficult to implement 
\cite{gustavsson} and does not detract from the scalability.
In conclusion, $Alq_{3}$ based quantum processors (1) are scalable, (2) are capable of fault-tolerant operation 
at room temperature, (3) possibly have a high degree of gate fidelity, and (4) lend themselves to an elegant qubit 
read out scheme. This makes them attractive candidates for quantum computers.

We acknowldege support by the US Air Force Office of Scientific Research under grant 
FA9550-04-1-0261 and by the US National Science Foundation under grants ECCS-0608854 and CCF-0726373.
The conclusions in this Letter are of the authors and do not represent the view of the Foundation.

\clearpage

\begin{figure}
\epsfig{file=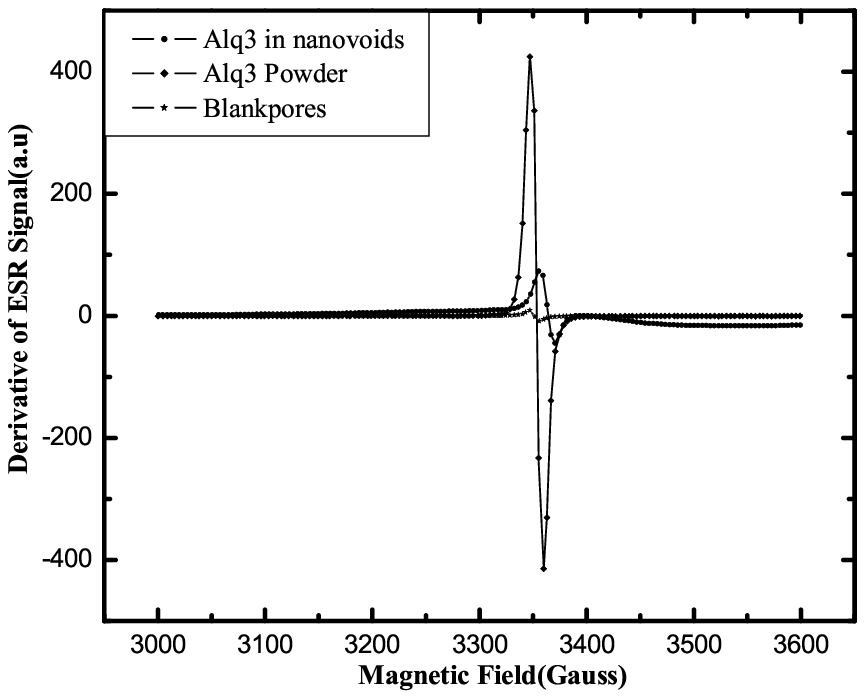} 
\caption{First derivative in magnetic field of the electron spin resonance spectrum 
corresponding to $g$ = 2. The three curves are the data for the blank alumina matrix, 
the $Alq_{3}$ powder and $Alq_{3}$ molecules in nanovoids. The temperature is 10 K.} 
\end{figure}

\begin{figure}
\epsfig{file=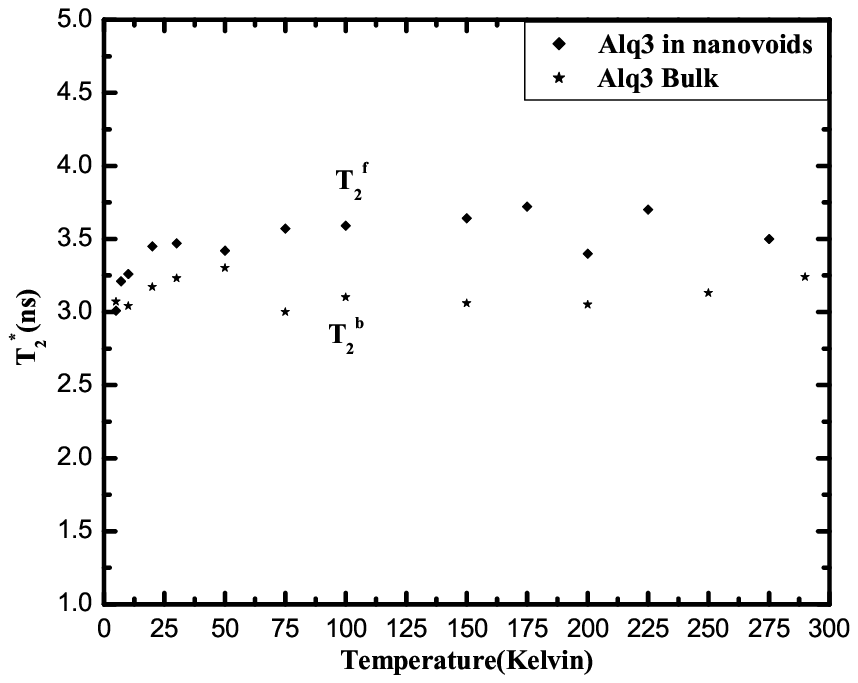}
\caption{ Transverse spin relaxation times as a function of temperature for $g$ = 2 resonance. 
The two plots are for bulk $Alq_{3}$ powder ($ T_2^{b}$) and few $Alq_{3}$ molecules  
in nanovoids ($T_2^{f}$).} 
\end{figure}

\begin{figure}
\epsfig{file=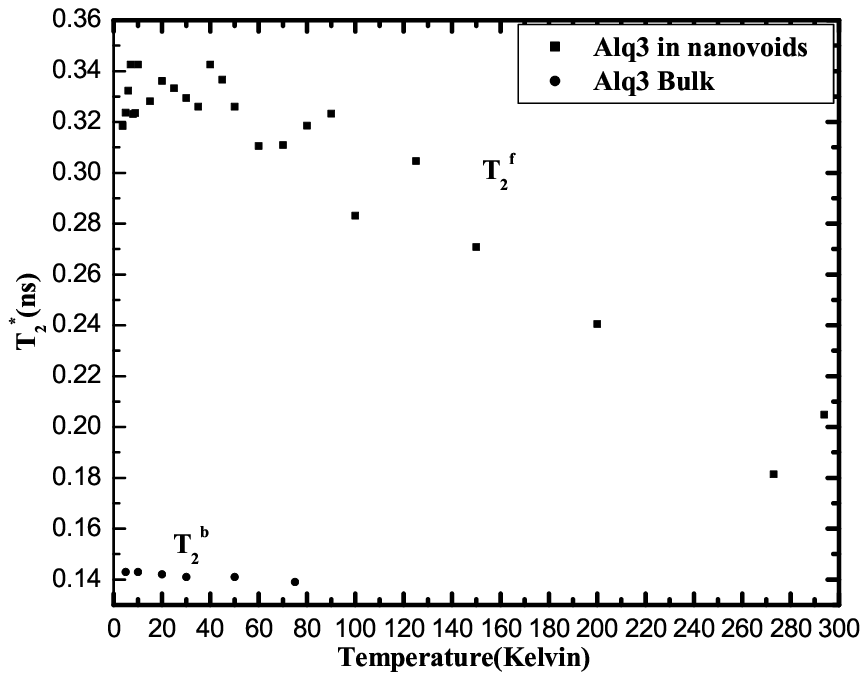}
\caption{ Transverse spin relaxation times as a function of temperature for $g$ = 4 resonance. 
The two plots are for bulk $Alq_{3}$ powder ($ T_2^{b}$) and few $Alq_{3}$ molecules  in nanovoids ($T_2^{f}$).} 

\end{figure}
\end{document}